\documentclass[12pt,reqno]{amsart}
\usepackage{amsfonts}
\usepackage{mathrsfs}
\usepackage{amsmath}
\usepackage{amssymb,amsfonts}

\newtheorem{thm}{Theorem}[section]
\newtheorem{lem}{Lemma}[section]

\numberwithin{equation}{section}

\def\pf{{\textit {Proof:} }}

\newcommand{\mysection}[1]{\section{#1}\setcounter{equation}{0}}

\newfont{\bb}{msbm10 at 11pt}

\newcommand{\bal}{\begin{aligned}}      \newcommand{\eal}{\end{aligned}}
\newcommand{\ba}{\begin{array}}      \newcommand{\ea}{\end{array}}
\newcommand{\bc}{\begin{center}}     \newcommand{\ec}{\end{center}}
\newcommand{\be}{\begin{enumerate}}  \newcommand{\ee}{\end{enumerate}}
\newcommand{\beq}{\begin{eqnarray}}  \newcommand{\eeq}{\end{eqnarray}}
\newcommand{\beQ}{\begin{eqnarray*}} \newcommand{\eeQ}{\end{eqnarray*}}
\newcommand{\bi}{\begin{itemize}}    \newcommand{\ei}{\end{itemize}}
\newcommand{\bt}{\begin{tabular}}    \newcommand{\et}{\end{tabular}}
\newcommand{\bdm}{\begin{displaymath}} \newcommand{\edm}{\end{displaymath}}
\newcommand{\pt}{\partial}

\newcommand{\pr}{\partial r}
\newcommand{\rw}{\rightarrow}

\newcommand{\Lrw}{\Longrightarrow}

\newcommand{\ep}{\epsilon}

\newcommand{\Ba}{\Big |}

\newcommand{\ur}{\left(1+u+r\right)}
\newcommand{\uu}{\left(1+u\right)}

\def\qed{\hfill{Q.E.D.}\smallskip}

\newcommand{\ls}{\setlength{\baselineskip}{12pt}
                 \setlength{\parskip}{3mm}}

\begin{document}

\allowdisplaybreaks

\title[Einstein-scalar-field equations]{Spherically symmetric Einstein-scalar-field equations with potential for wave-like decaying null infinity}

\author[C Liu]{Chuxiao Liu$^{1,4}$}
\author[X Zhang]{Xiao Zhang$^{2,3,4}$}

\address[]{$^{1}$School of Mathematics and Information Science, Guangxi University, Nanning, Guangxi 530004, PR China}
\address[]{$^{2}$Academy of Mathematics and Systems Science, Chinese Academy of Sciences, Beijing 100190, PR China}
\address[]{$^{3}$School of Mathematical Sciences, University of Chinese Academy of Sciences, Beijing 100049, PR China}
\address[]{$^{4}$Guangxi Center for Mathematical Research, Guangxi University, Nanning, Guangxi 530004, PR China}

\email{cxliu@gxu.edu.cn$^{1,4}$}
\email{xzhang@amss.ac.cn$^{2,3}$, xzhang@gxu.edu.cn$^{4}$}

\subjclass[2000]{53C50, 58J45, 83C05}
\keywords{Einstein-scalar-field equations with potential; spherically symmetric Bondi-Sachs metrics; null infinity; wave-like decaying conditions}

\date{}

\begin{abstract}

We prove the global existence and uniqueness of classical solutions with small initial data and with wake-like decaying null infinity for the spherically symmetric Einstein-scalar-field equations with potential, where the scalar potential $V$ satisfies \eqref{V} and \eqref{V1}.
\end{abstract}

\maketitle \pagenumbering{arabic}

\mysection{Introduction}
\ls
\subsection{Spherically symmetric Einstein-scalar field equations with potential}
\ls

We provide some well-known facts about the spherically symmetric Einstein-scalar field equations with potential; for details, see, e.g. \cite{C1, WFAG}.
In general relativity, metrics of spherically symmetric spacetime can be written as
\beq
ds^2 =-g qdu^2-2gdudr+r ^2 \left(d \theta ^2 +\sin ^2 \theta d \psi ^2\right)\label{metric}
\eeq
in Bondi coordinates, where $g(u, r)$ and $q(u,r)$ are $C^2$ and nonnegative over $(0, \infty)$.
Denote by $D$ the derivative along the incoming light rays
\beQ
D=\frac{\pt}{\pt u}-\frac{q}{2}\frac{\pt}{\pt r}.%\label{a-3}
\eeQ
The null frames are given by
\beQ
    \vec{n}=\frac{1}{\sqrt{g q}} D, \quad \vec{l}=\frac{1}{\sqrt{g q^{-1}}} \frac{\partial }{\partial r}, \quad
    e_1=\frac{1}{r}\frac{\partial }{\partial \theta}, \quad
    e_2=\frac{1}{r \sin \theta }\frac{\partial }{\partial \psi}.
    \eeQ
Let $T_{\mu\nu}$ be a symmetric 2-tensor which is spherically symmetric and divergence-free, i.e., it satisfies that
\beQ
\nabla ^\mu T_{\mu\nu}=0.
\eeQ
Define
\begin{align*}
G_{\mu\nu}=R_{\mu\nu}-\frac{R}{2} g_{\mu\nu},\quad
E_{\mu\nu}=G_{\mu\nu}-8\pi T_{\mu\nu}.
\end{align*}
For spherically symmetric metrics, the only nonzero components are
\beQ
E(\vec{n}, \vec{l}), \,\,E(\vec{l}, \vec{l}), \,\,E(\vec{n}, \vec{n}), \,\,E(e_1, e_1), \,\,E(e_2, e_2).
\eeQ

By the twice contacted Bianchi identities, $E_{ij}$ is divergence-free. This implies that if
\beq
E(\vec{n}, \vec{l})=E(\vec{l}, \vec{l})=0, \label{2E}
\eeq
then
\begin{align*}
E(e_1, e_1)=E(e_2, e_2)=0,\quad \frac{\pt}{\pt r}\Big(E(\vec{n},\vec{n})gqr^2\Big)=0,
\end{align*}
where
\begin{align}
E(\vec{n},\vec{n})gqr^2=
-\frac{\partial \ln g}{\partial u} r q +\frac{\partial q}{\partial u} r
+\frac{1}{2} \frac{\partial \ln g}{\partial r} r q^2 -8\pi r^2 T(D,D). \label{enn}
\end{align}

We introduce the first regularity condition at $r=0$.

\noindent{\bf Regularity Condition I}: For each $u$,
\beq
\lim _{r \rightarrow 0} \Big(E(\vec{n},\vec{n})g(u,r)q(u,r)r^2\Big)=0. \label{regI}
\eeq
Under (\ref{regI}), the Einstein field equations with the energy-momentum tensor $T_{ij}$ are equivalent to (\ref{2E}).

Assume that the energy-momentum tensor is given by
\beq
T_{\mu\nu}=\pt_{\mu}\phi\pt_{\nu}\phi-\frac{1}{2}g_{\mu\nu}g^{\alpha\beta}
\pt_{\alpha}\phi\pt_{\beta}\phi+g_{\mu\nu}V(\phi),\label{a-8}
\eeq
where $\phi(u,r)$ is a real, $C^2$ scalar field over $(0, \infty)$, and $V(\phi)$ is the scalar potential. The divergence-free condition of $T_{\mu\nu}$ gives that
\beq
\Box\phi=-\frac{\pt V(\phi)}{\pt \phi}.\label{a-9}
\eeq

Throughout the paper, we denote by
\beQ
\bar f(u,r)=\frac{1}{r}\int_{0}^{r}{f(u,r')dr'}
\eeQ
the integral average of $f(u,r)$ with respect to $r$. Define
\beq
h=\frac{\pt(r\phi)}{\pt r}.    \label{h}
\eeq
It is clear that
\begin{align}
E(\vec{n}, \vec{l})=0 \Lrw &  \frac{\partial (rq)}{\partial r}=g+8\pi gr^2V(\phi),\label{n-l}\\
E(\vec{l}, \vec{l})=0 \Lrw & \frac{\partial \ln g}{\partial r}=4\pi r\Big(\frac{\partial\phi}{\partial r}\Big)^2.\label{l-l}
\end{align}

We introduce the second regularity condition at $r=0$ as well as the boundary condition at $r=\infty$.

\noindent{\bf Regularity Condition II}: For each $u$,
\begin{align}
\lim _{r \rightarrow 0} \Big(r \phi(u,r) \Big)=\lim _{r \rightarrow 0} \Big(r q(u,r) \Big)=0. \label{regII}
\end{align}

\noindent{\bf Boundary Condition}: For each $u$,
\beq
\lim _{r \rightarrow \infty} g(u,r)=\lim _{r \rightarrow \infty} q(u,r)=1. \label{bdy}
\eeq
Under (\ref{regII}), (\ref{h}) and (\ref{n-l}) give
\begin{align*}
\phi =\bar h,\quad q=\bar g +\frac{8\pi}{r} \int_0 ^r g s^2 V(\bar h) ds.
\end{align*}
Under (\ref{bdy}), (\ref{metric}) is asymptotically flat, and (\ref{l-l}) gives
\beQ
g=\exp\left\{-4\pi\int_{r}^{\infty}{(h-\bar h)^2\frac{ds}{s}}\right\}.
\eeQ
Therefore, under the regularity conditions (\ref{regI}), (\ref{regII}), and the boundary condition (\ref{bdy}), the Einstein-scalar field equations with scalar potential $V$ are equivalent to
\beq
\begin{matrix}
\left\{
\begin{aligned}
g&=\exp\left\{-4\pi\int_{r}^{\infty}{(h-\bar{h})^2\frac{ds}{s}}\right\},\\
q&=\bar g +\frac{8\pi}{r} \int_0 ^r g s^2 V(\bar h) ds,\\
D h&=\frac{1}{2r}(g-q)(h-\bar{h})+4\pi gr(h-\bar h)V(\bar h)+\frac{gr}{2}\frac{\pt V(\bar h)}{\pt \bar h}.  \label{e-k-g-2}
\end{aligned}\right.
\end{matrix}
\eeq

The Bondi mass $M_B(u)$ for each $u$ and the final Bondi mass $M_{B1}$ are given by \cite{BBM, LZ}
\beQ
M_B(u)=\lim _{r \rightarrow \infty} \frac{r}{2}\Big(1-q\Big), \quad M _{B1}=\lim _{u \rightarrow \infty} M_B(u).
\eeQ
The Bondi-Christodoulou mass $M(u)$ for each $u$ and the final Bondi-Christodoulou mass $M_1$ are given by \cite{C1, LZ}
\beQ
M(u)=\lim _{r \rightarrow \infty} \frac{r}{2}\Big(1-\frac{q}{g}\Big), \quad M _1=\lim _{u \rightarrow \infty} M(u).
\eeQ

\subsection{Main results}

$(i)\,\,V=0$. In this case, Christodoulou first studied the global existence and uniqueness of classical solutions with small initial data, and of generalized solutions with large initial data, for particle-like decaying null infinity \cite{C1, C2, C3}. In particular, he proved that the solution satisfies the following uniformly decaying estimates
\beQ
\big| h(u,r) \big| \leq \frac{C}{(1+u+r)^{3}}, \quad \Big| \frac{\partial h}{\partial r}(u,r)\Big| \leq \frac{C}{(1+u+r)^{4}},
\eeQ
and the corresponding spacetime is future causally geodesically complete with vanishing final Bondi-Christodoulou mass.

Christodoulou also showed the unique spherically symmetric global solution exists for the characteristic initial-value problem for small bounded variation norms under the double null coordinates \cite{C4}
\beQ
ds^2=-\Omega ^2 du dv +r^2 \big(d\theta^2+\sin^2\theta d\phi^2 \big).
\eeQ
Later, his result was extended to the more general situation by Luk, Oh and Yang, and they showed the unique spherically symmetric global solution exists and
the resulting spacetime is future causally geodesically complete if the initial data satisfies
\beQ
\begin{aligned}
\int _u ^v \big| \Phi(v') \big| dv'  \leq \epsilon \big(v-u \big)^{1-\gamma}, \quad
|\Phi(v)| + \Big| \frac{\partial \Phi} {\partial v} (v)\Big| \leq \epsilon
\end{aligned}
\eeQ
where $\gamma$, $\epsilon$ are certain positive constants. However, the solution does not have uniformly decaying estimates unless the initial data satisfies
\beQ
\sup\limits_{v\in[u_0,\infty)}{\bigg\{(1+v)^\ep |\Phi(v)|+(1+v)^{\ep +1}|\pt_v\Phi(v)|\bigg\}}\leq A_0
\eeQ
further for some $A_0>0$ and $\ep>1$ \cite{LO, LOY}.

In \cite{LZ}, Liu and Zhang studied the global existence and uniqueness of classical solutions with small initial data, and of generalized solutions with large initial data, for wave-like decaying null infinity. They proved that the solution satisfies the following uniformly decaying estimates
\beQ
|h(u,r)|\leq \frac{C}{(1+u+r)^{1+\ep}},\quad \Ba\frac{\pt h}{\pr}(u,r)\Ba\leq\frac{C}{(1+u+r)^{1+\ep}}
\eeQ
for $0<\ep \leq 2$, and the corresponding spacetime is future causally geodesically complete with vanishing final Bondi mass.

$(ii)\,\,V \neq 0$. In \cite{C01}, Chae proved the global existence and uniqueness of the spherically symmetric Einstein-(nonlinear)Klein-Gordon system for small initial data with scalar potential
\beQ
V(\phi)=-\frac{1}{p+1}|\phi|^{p+1}
\eeQ
for particle-like decaying null infinity, where $p \geq k$, $k=3$, $4$. He proved that the solution satisfies the following uniformly decaying estimates
\beQ
|h(u,r)|\leq \frac{C}{(1+u+r)^{k-1}},\quad \left|\frac{\pt h}{\pt r}(u,r)\right|\leq \frac{C}{(1+u+r)^{k}}.
\eeQ
It was extended to the more general situation by Wijayanto, Fadhilla, Akbar and Gunara \cite{WFAG} where the scalar potential satisfies
\begin{align}
|V(\phi)|+\left|\frac{\pt V(\phi)}{\pt \phi}\right||\phi|+\left|\frac{\pt^2 V(\phi)}{\pt \phi^2}\right||\phi|^2\leq K_0|\phi|^{p+1},\label{V}
\end{align}
where $p \geq k\geq 3$, constant $K_0 > 0$. (The complex version of \eqref{V} was already used for the Einstein-Maxwell-Higgs system \cite{C02}.)

In this paper, we prove the following theorem for wave-like decaying null infinity.
\begin{thm}\label{M}
Let $\ep\in(0,2]$. Given initial data $h(0,r)\in C^1[0,\infty)$, denote
\beQ
d=\sup\limits_{r\geq 0}{\left\{(1+r)^{1+\ep}\left(|h(0,r)|+\left|\frac{\pt h}{\pt r}(0,r)\right|\right)\right\}}.
\eeQ
Then there exists $\delta>0$ such that if $d<\delta$, there exists a unique global classical solution $$h(u,r)\in C^1([0,\infty)\times[0,\infty))$$ of (\ref{e-k-g-2}) with $h(0,r)$ as the initial data for scalar potential $V(\phi)$ satisfying \eqref{V} where
\begin{align}
\left\{\begin{aligned}
            &p \geq 2+\ep,\quad   &\sqrt{2}-1< \ep \leq 2,\\
            &p> \frac{3+\ep}{1+\ep}, & \sqrt{2}-1 \geq \ep>0,\\
\end{aligned}
\right.\label{V1}
\end{align}
and the solution satisfies the following uniformly decaying estimates
\beQ
|h(u,r)|\leq \frac{C}{(1+u+r)^{1+\ep}},\quad \left|\frac{\pt h}{\pt r}(u,r)\right|\leq \frac{C}{(1+u+r)^{1+\ep}}.
\eeQ
Moreover, the final Bondi mass vanishes and the corresponding spacetime is future causally geodesically complete.
\end{thm}

For $\ep$, $p$ satisfying the assumption of Theorem \ref{M}, we define
\begin{align*}
\omega=\min{\{p,3\}}>2, \quad \mu=\min\left\{1+\ep,\,\,(\omega-1)\ep+\omega-2\right\}>1
\end{align*}
throughout the paper. It is straightforward that
\begin{align*}
(p-1)\ep+p-2>1\Longrightarrow 1<(\omega-1)\ep+(\omega-2)\leq 2\ep+1.
\end{align*}

We adopt the main argument of \cite{LZ} to derive the wave-like decaying estimates with nontrivial scalar potential in order to prove the theorem. It can be easily followed from \cite{WFAG} when $p \geq 3$. However, it needs much careful and nontrivial analysis to achieve it when $p<3$.

We refer to \cite{FWY, K, KWY, ML, LR, LO1, W} and the references therein for existence and uniqueness of Einstein fields equations coupled with scalar, Maxwell and Klein-Gordon fields on non-spherically symmetric metrics for non-wave-like decaying null infinity.

The paper is organized as follows: In Section 2, we derive the main estimates for the solution of the spherically symmetric Einstein-scalar equations with scalar potential. In Section 3, we prove the contraction mapping property for wave-like decaying solutions. In Section 4, we prove the main theorem.

\mysection{Main estimates}
\ls

In this section, we follow the arguments in \cite{C01, C02, LZ, WFAG} to derive some key estimates. Denote $X$ the space of all $C^1$ function $h(u,r)$
defined on $[0,\infty)\times[0,\infty)$ such that the following norm
\begin{align*}
\|h\|_{X}=\sup\limits_{u\geq 0}\sup\limits_{r\geq 0}{\left\{(1+u+r)^{1+\ep}\left(|h(u,r)|+\left|\frac{\pt h}{\pt r}(u,r)\right|\right)\right\}}
\end{align*}
is finite. Let
\begin{align*}
B(x)=\Big\{f\in X \,\Big|\,\|f\|_X\leq x\Big\}
\end{align*}
be the closed ball of radius $x$ in $X$. As in \cite{WFAG}, consider the mapping
\begin{align*}
h\mapsto \mathfrak{F}(h)
\end{align*}
which is defined as the solution of the following equation
\beq
\begin{aligned}
D\mathfrak{F}=&\frac{1}{2r}(\mathfrak{F}-\bar h)(g-q)+4\pi gr(\mathfrak{F}-\bar h)V+\frac{gr}{2}\frac{\pt V}{\pt \bar h}\\
             =&\left(\frac{g-q}{2r}+4\pi grV\right)\mathfrak{F}
              -\left(\frac{g-q}{2r}+4\pi grV\right)\bar h+\frac{gr}{2}\frac{\pt V}{\pt\bar h}.\label{l-1}
\end{aligned}
\eeq
with the initial data
\beQ
\mathfrak{F}(0,r)=h(0,r).
\eeQ
Let $r(u)=\chi(u;r_0)$ be the characteristic which satisfies the ordinary differential equation
\beQ
\frac{dr}{du}=-\frac{q(u,r)}{2},\quad r(0)=r_0.
\eeQ
Denote $r_1=\chi(u_1;r_0)$. Integrating (\ref{l-1}) along the characteristic $\chi$, we can obtain the explicit expression of $\mathfrak{F}$
\beq
\begin{aligned}
\mathfrak{F}(u_1,r_1)=& h(0,r_0)\exp\left\{\int_{0}^{u_1}{\left[\frac{g-q}{2r}+4\pi grV\right]_{\chi}du}\right\}\\
&+\int_{0}^{u_1}{\exp\left\{\int_{u}^{u_1}{\left[\frac{g-q}{2r}+4\pi grV\right]_{\chi}du'}\right\}[f]_{\chi}du}.
\end{aligned}\label{int-f}
\eeq

\begin{lem}\label{l1}
Let $\ep\in(0,2]$. Given the initial data $h(0,r)\in C^1[0,\infty)$ such that
\begin{align*}
d=\|h(0,r)\|_{X}=\sup\limits_{r\geq 0}{\left\{(1+r)^{1+\ep}\left(|h(0,r)|+\left|\frac{\pt h}{\pt r}(0,r)\right|\right)\right\}}.
\end{align*}
Assume
\begin{align*}
\|h(u,r)\|_X=x.
\end{align*}
Then the solution of (\ref{l-1}) satisfies
\begin{align}
|\mathfrak{F}(h)(u,r)|\leq &\frac{C(d+x^3+x^p+x^{p+2})\exp\left[C(x^2+x^{p+1})\right]}{(1+u+r)^{1+\ep}}, \label{l1-a}\\
\left|\frac{\pt \mathfrak{F}(h)}{\pt r}(u,r)\right|\leq & \frac{C(d+x^3+x^p+x^{p+2})\exp\left[C(x^2+x^{p+1})\right]}{(1+u+r)^{1+\ep}} P(x). \label{l1-b}
\end{align}
Moreover, there exists $x_0>0$ such that for any $x\in (0,x_0)$, $d\leq F_1(x)$,
\begin{align*}
\mathfrak{F}(B(x))\subset B(x).
\end{align*}
Here
\begin{align*}
P(x)=&1+x^2+x^{p+1}+x^{p+3},\\
F_1(x)=&\frac{x\exp\left[-A(x^2+x^{p+1})\right]}{A(1+x^2+x^{p+1}+x^{p+3})}-(x^3+x^p+x^{p+2}),
\end{align*}
and $A$ and $C$ are some positive constants.
\end{lem}
\pf Let $c=\frac{6}{\ep|1-\ep|}$ for $\ep\neq 1$ and $c=24$ for $\ep=1$. In \cite{LZ}, we proved that if
\begin{align*}
|h(u,r)|\leq \frac{x}{\ur^{1+\ep}},\quad \left|\frac{\pt h}{\pt r}(u,r)\right|\leq \frac{x}{\ur^{1+\ep}}
\end{align*}
for some constant $x>0$, then the following estimates hold
\begin{align}
|\bar h|&\leq \frac{2x}{\ep}\frac{1}{\uu^{\ep}\ur},\label{l-2a}\\
|h-\bar h|&\leq \frac{cxr\uu^{1-\ep}}{\ur^2},\label{l-3}\\
|g-\bar g|&\leq \frac{4\pi c^2x^2}{3}\frac{r^2}{\uu^{2\ep-1}\ur^3},\label{l-5}\\
\bar g(u,r)&\geq \exp\left(-\frac{2\pi c^2x^2}{3}\right).
\end{align}

These estimates yield the following claim.

{\em Claim}: Let $C_1=\frac{2^{p+2}K_0}{\omega(\omega-1)(\omega-2)\ep^{p+1}}$. Then, for $p>2$, it holds that
\begin{align}
\int_{0}^{r}{gs^2V(\bar h)ds}\leq \frac{C_1x^{p+1}r^3}{\uu^{(\omega+1)\ep+1}\ur^\omega}.\label{l-7}
\end{align}

{\em Proof of Claim:} If $2<p<3$, using (\ref{l-2a}), we obtain
\beQ
\begin{aligned}
\int_{0}^{r}{gs^2V(\bar h)ds}&\leq \frac{K_02^{p+1}x^{p+1}}{\ep^{p+1}\uu^{(p+1)\ep}}\int_{0}^{r}{
\frac{s^2}{(1+u+s)^{p+1}}ds}\\
&\leq \frac{K_02^{p+1}x^{p+1}\left[F(r)+2\uu^{p-1}r^3\right]}{p(p-1)(p-2)\uu^{(p+1)\ep+p}\ur^p},\\
\end{aligned}
\eeQ
where
\begin{align*}
F(r)=&2\uu^2\ur^p-p(p-1)\uu^p\ur^2\\
&+2p(p-2)\uu^{p+1}\ur \\
&-(p-1)(p-2)\uu^{p+2}-2\uu^{p-1}r^3.
\end{align*}
It is straightforward that
\begin{align*}
F^{(4)}(r)\leq 0.
\end{align*}
By analyzing the monotonicity, we obtain
\beQ
\begin{aligned}
&  F'''(r)\leq F'''(0)=\big[2p(p-1)(p-2)-12 \big]\uu^{p-1}\leq 0\\
\Longrightarrow & F''(r)\leq F''(0)=0
\Longrightarrow F'(r)\leq F'(0)=0
\Longrightarrow F(r)\leq F(0)=0.
\end{aligned}
\eeQ
This implies \eqref{l-7}.

If $p\geq 3$, we can use a similar argument to prove that
\begin{align*}
\int_{0}^{r}{gs^2V(\bar h)ds}&\leq \frac{K_02^{p+1}x^{p+1}}{\ep^{p+1}\uu^{4\ep}}\int_{0}^{r}{\frac{s^2}{(1+u+s)^4}ds}\\
&= \frac{K_02^{p+1}x^{p+1}r^3}{3\ep^{p+1}\uu^{4\ep+1}\ur^3}.
\end{align*}
Therefore, \eqref{l-7} follows.

Let $C_2=\frac{4\pi c^2}{3}+8\pi C_1 $. We have
\beq
\begin{aligned}
|g-q|\leq & |g-\bar g|+\frac{8\pi}{r}\int_{0}^{r}{gs^2V(\bar h)ds}\\
\leq & \frac{C_2r^2(x^2+x^{p+1})}{\uu^{2\ep+2-\omega}\ur^{\omega}}.\label{l-8}
\end{aligned}
\eeq
Denote
\beQ
k(x)=\exp\left(-\frac{2\pi c^2x^2}{3}\right)- 8\pi C_1x^{p+1}, \quad x \geq 0.
\eeQ
Clearly, $k(x)$ is monotonically decreasing and
\beQ
k(0)=1,\quad k(\infty)=-\infty.
\eeQ
Then there exists $x_1>0$ such that for any $x\in [0,x_1)$,
\beQ
0<k(x)\leq 1.
\eeQ
Thus, on $[0,x_1)$,
\begin{align}
q \geq \bar g(u,0)-\left|\frac{8\pi}{r}\int_{0}^{r} {gs^2V(\bar h)ds}\right| \geq k(x)>0. \label{qq}
\end{align}
Therefore
\begin{align*}
r(u)=r_1+\frac{1}{2}\int_{u}^{u_1}{q(u,\chi(u;r_0))du}\geq r_1+\frac{k}{2}(u_1-u).
\end{align*}
This yields
\beQ
1+r(u)+u\geq 1+r_1+\frac{k}{2}(u_1-u)+u\geq\frac{k}{2}(1+r_1+u_1).%\label{l-11}
\eeQ
Taking $u=0$, we obtain
\beq
|h(0,r_0)|\leq \frac{d}{(1+r_0)^{1+\ep}}\leq \frac{2^{1+\ep}d}{k^{1+\ep}(1+r_1+u_1)^{1+\ep}}.\label{l-12}
\eeq
Let $C_3=\frac{C_2}{2\ep}+\frac{2^{p+3}K_0\pi}{\ep^{p+2}}$. Using (\ref{l-8}), we obtain
\beq
\begin{aligned}
\int_{0}^{u_1}&{\left[\frac{1}{2r}\left|(g-q)\right|+4\pi gr|V(\bar h)|\right]_{\chi}du}\\
\leq &\int_{0}^{u_1}\left[\frac{C_2r^2(x^2+x^{p+1})}{2r\uu^{\ep+1}\ur^{1+\ep}}\right.\\
&\left.+\frac{4\pi rK_02^{p+1}x^{p+1}}{\ep^{p+1}\uu^{(p+1)\ep}\ur^{p+1}}\right]_{\chi}du\\
\leq &C_3(x^2+x^{p+1}).\label{l-13}
\end{aligned}
\eeq

Now we derive \eqref{l1-a}. Using (\ref{l-5}) and (\ref{l-8}), we have
\begin{align*}
\left|-\frac{1}{2r}(g-q)\bar h\right|
\leq &\frac{C_2r^2(x^2+x^{p+1})}{2r\uu^{\ep+1}\ur^{1+\ep}}\frac{2x}{\ep\uu^{\ep}\ur}\\
\leq &\frac{C_2(x^3+x^{p+2})}{\ep\uu^{2\ep+1}\ur^{1+\ep}},\\
|4\pi gr V(\bar h)\bar h|
\leq& \frac{4\pi rK_02^{p+1}x^{p+1}}{\ep^{p+1}\uu^{(p+1)\ep}\ur^{p+1}}\frac{2x}{\ep\uu^{\ep}\ur}\\
\leq& \frac{2^{p+4}\pi K_0x^{p+2}}{\ep^{p+2}\uu^{2\ep+1}\ur^{1+\ep}},\\
\left|\frac{gr}{2}\frac{\pt V(\bar h)}{\pt \bar h}\right|
\leq &\frac{2^pK_0x^pr}{2\ep^p\uu^{p\ep}\ur^p}\leq\frac{2^{p-1}K_0x^p}{\ep^p\uu^{p\ep}\ur^{p-1}}.
\end{align*}
Let $C_4=\max{\left\{\frac{2^{p-1}K_0}{\ep^p},\,\,2C_3\right\}}$, and
\begin{align*}
f=-\left(\frac{g-q}{2r}+4\pi grV\right)\bar h+\frac{gr}{2}\frac{\pt V}{\pt\bar h}.
\end{align*}
We obtain
\beQ
\begin{aligned}
|f|&\leq \left|-\frac{1}{2r}(g-q)\bar h\right|+|4\pi gr V(\bar h)\bar h|+\left|\frac{gr}{2}\frac{\pt V(\bar h)}{\pt \bar h}\right|\\
&\leq \frac{C_4(x^3+x^p+x^{p+2})}{\uu^{(\omega-1)\ep+\omega-2}\ur^{1+\ep}}.
\end{aligned}
\eeQ
This implies that
\beQ
\begin{aligned}
\int_{0}^{u_1}{[f]_{\chi}du}&\leq\int_{0}^{u_1}{\left[\frac{C_4(x^3+x^p+x^{p+2})}{\uu^{(\omega-1)\ep+\omega-2}\ur^{1+\ep}}\right]_{\chi}du}\\
&\leq \frac{2^{1+\ep}C_4(x^3+x^p+x^{p+2})}{k^{1+\ep}(1+r_1+u_1)^{1+\ep}}\int_{0}^{u_1}{\frac{1}{\uu^{(\omega-1)\ep+\omega-2}}du}\\
&\leq \frac{2^{1+\ep}C_4(x^3+x^p+x^{p+2})}{[(\omega-1)\ep+\omega-3] k^{1+\ep}(1+r_1+u_1)^{1+\ep}}.\\
\end{aligned}
%\label{l-15}
\eeQ
Let  $C_5=\frac{1}{k^{1+\ep}}\max{\left\{2^{1+\ep},\,\,\frac{C_42^{1+\ep}}{(\omega-1)\ep+\omega-3}\right\}}$. Using \eqref{int-f} and the above estimates, we can obtain (\ref{l1-a}).

Next, we derive (\ref{l1-b}). Define
\begin{align*}
\mathfrak{G}(u,r)=\frac{\pt \mathfrak{F}}{\pt r}(u,r), \quad \mathfrak{G}(0,r_0)=\frac{\pt h}{\pt r}(0,r_0).
\end{align*}
Differentiate (\ref{e-k-g-2}) with respect to $r$, we have (cf. (3.36) in \cite{WFAG})
\begin{align}
D\mathfrak{G}=f_1\mathfrak{G}+f_2,\label{l-19}
\end{align}
where
\begin{align*}
f_1=&\frac{1}{2}\frac{\pt q}{\pt r}+\frac{1}{2r}(g-q)+4\pi grV(\bar h),\\
f_2=&\left[\frac{1}{2r}\frac{\pt}{\pt r}(g-q)-\frac{1}{2r^2}(g-q)+\frac{\pt g}{\pt r}4\pi rV(\bar h)\right.\\
&\left.+4\pi gV(\bar h)+4\pi gr\frac{\pt V(\bar h)}{\pt \bar h}\frac{\pt \bar h}{\pt r}\right](\mathfrak{F}-\bar h)\\
&-\left[\frac{1}{2r}(g-q)+4\pi grV(\bar h)-\frac{gr}{2}\frac{\pt^2V(\bar h)}{\pt \bar h^2}\right]\frac{\pt \bar h}{\pt r}\\
&+\left[\frac{\pt g}{\pt r}\frac{r}{2}+\frac{g}{2}\right]\frac{\pt V(\bar h)}{\pt \bar h}.
\end{align*}
Integrating (\ref{l-19}) along the characteristic $\chi$, we obtain
\beq
\begin{aligned}
\mathfrak{G}(u_1,r_1)=&\mathfrak{G}(0,r_0)\exp{\left\{\int_{0}^{u_1}{[f_1]_{\chi}du}\right\}}\\
&+\int_{0}^{u_1}{\exp{\left\{\int_{u}^{u_1}{[f_1]_{\chi}du'}\right\}}[f_2]_{\chi}du}.\\
\end{aligned}\label{int-g}
\eeq
Let $C_6=\max{\left\{\frac{4\pi c^2}{3},\,\,[8+4\omega(\omega-1)(\omega-2)]\pi C_1\right\}}$. Using the first and the second equations of \eqref{e-k-g-2},   (\ref{l-3}), (\ref{l-5}), and (\ref{l-7}), we have
\begin{align}
\left|\frac{\pt g}{\pt r}\right|&\leq |g|\frac{4\pi}{r}(h-\bar h)^2\leq \frac{4\pi c^2x^2r}{\uu^{2\ep+2-\omega}\ur^\omega}, \label{l-23}\\
\left|\frac{\pt q}{\pt r}\right|&\leq \frac{g-\bar g}{r}+\frac{8\pi}{r^2}\int_{0}^{r}{gs^2|V(\bar h)|ds}+8\pi gr|V(\bar h)|   \nonumber\\
&\leq \frac{C_6(x^2+x^{p+1})r}{\uu^{2\ep+2-\omega}\ur^{\omega}}. \label{l-22}
\end{align}
Let $C_7=\max{\left\{\frac{C_2+C_6}{2}+2\pi c^2+\frac{2^{p+3}\pi K_0}{\ep^{p+1}}+\frac{2^{p+2}K_0\pi c}{\ep^p},\,\frac{2^{p+5}\pi^2c^2K_0}{\ep^{p+1}}\right\}}$, $C_8=\max{\left\{\frac{2^{p-2}K_0(2+c\ep)}{\ep^p},\,\frac{cC_2}{2}
+\frac{2^{p+1}K_0\pi c(4+c\ep)}{\ep^{p+1}}\right\}}$. Using (\ref{V}), (\ref{l-3}), (\ref{l-23}), and (\ref{l-22}), we have
\begin{align*}
&\bigg|\frac{1}{2r}\frac{\pt}{\pt r}(g-q)-\frac{1}{2r^2}(g-q)+\frac{\pt g}{\pt r}4\pi rV(\bar h)+4\pi gV(\bar h)\\
&+4\pi gr\frac{\pt V(\bar h)}{\pt \bar h}\frac{\pt \bar h}{\pt r}\bigg| \leq \frac{C_7(x^2+x^{p+1}+x^{p+3})}{\uu^{\mu}\ur^{1+\ep}},\\
&\bigg|\frac{1}{2r}(g-q)+4\pi grV(\bar h)-\frac{gr}{2}\frac{\pt^2V(\bar h)}{\pt \bar h^2}\bigg|\left|\frac{\pt \bar h}{\pt r}\right|\\
&+\bigg|\frac{\pt g}{\pt r}\frac{r}{2}+\frac{g}{2}\bigg|\left|\frac{\pt V(\bar h)}{\pt \bar h}\right| \leq \frac{C_8(x^3+x^p+x^{p+2})}{\uu^{\mu}\ur^{1+\ep}}.
\end{align*}
These together with (\ref{int-f}) yield
\beq
\begin{aligned}
|f_2|\leq & \frac{C_7(x^2+x^{p+1}+x^{p+3})}{\uu^{\mu}\ur^{1+\ep}}|\mathfrak{F}|\\
&+\frac{C_7(x^2+x^{p+1}+x^{p+3})}{\uu^{\mu}\ur^{1+\ep}}\cdot \frac{2x}{\ep\uu^{\ep}\ur}\\
&+\frac{C_8(x^3+x^p+x^{p+2})}{\uu^{\mu}\ur^{1+\ep}}.
\end{aligned}
\label{l-26}
\eeq
Let $C_9=\frac{1}{2\ep}(C_2+C_6+\frac{2^{p+4}\pi K_0}{\ep^{p+1}})$. Similar to (\ref{l-12}) and (\ref{l-13}), we have
\begin{align}
\left|\frac{\pt h}{\pt r}(0,r_0)\right|\leq &\frac{\|h(0,r_0)\|_{X}}{(1+r_0)^{1+\ep}}\leq \frac{2^{1+\ep}d}{k^{1+\ep}(1+u_1+r_1)^{1+\ep}},\label{l-27}\\
\int_{0}^{u_1}{|f_1|_{\chi}du}\leq & C_9(x^2+x^{p+1}).\label{l-28}
\end{align}
Let $C_{10}=\frac{1}{(\mu-1)k^{1+\ep}}\max{\left\{2^{1+\ep},C_3+C_9,C_5C_7,\frac{2C_7}{\ep},C_8\right\}}$. Using \eqref{int-g}, (\ref{l-26}), (\ref{l-27}), and (\ref{l-28}), we can obtain (\ref{l1-b}).

Finally, let $A=\max{\left\{C_3,C_5,C_{10}\right\}}$. Using (\ref{l1-a}), (\ref{l1-b}), we have
\begin{align*}
\|\mathfrak{F}\|_{X}\leq& C_5(d+x^3+x^p+x^{p+2})\exp\left[C_3(x^2+x^{p+1})\right]\\
&+C_{10}(d+x^3+x^p+x^{p+2})\exp\left[C_{10}(x^2+x^{p+1})\right]P(x)\\
\leq &A (d+x^3+x^p+x^{p+2})\exp\left[A(x^2+x^{p+1})\right] P(x).
\end{align*}
Define
\beQ
F_1(x)=\frac{x\exp\left[-A(x^2+x^{p+1})\right]}{A(1+x^2+x^{p+1}+x^{p+3})}-(x^3+x^p+x^{p+2}).
\eeQ
Obviously,
\beQ
F_1(0)=0,\quad F_1'(0)>0.
\eeQ
Thus, there exists $x_0\in (0,x_1)$ such that $F_1(x)$ is monotonically increasing on $[0,x_0]$ and achieves its maximum at point $x_0$. Then for any $x\in (0,x_0)$, if $d\leq F_1(x)$, we have
\begin{align*}
\|\mathfrak{F}\|_{X}\leq x \Longrightarrow \mathfrak{F}(B(x))\subset B(x).
\end{align*}
Therefore the proof of lemma is complete. \qed

\mysection{Contraction mapping property}
\ls

In this section, we show that $h\mapsto \mathfrak{F}(h)$ is a contraction mapping in $Y$, where $Y$ is the space of all $C^1$ function $h(u,r)$
defined on $[0,\infty)\times[0,\infty)$ with fixed initial data $h_0(r)$ such that the following norm
\begin{align*}
\|h\|_{Y}=\sup\limits_{u\geq 0}\sup\limits_{r\geq 0}{\left\{(1+u+r)^{1+\ep}|h(u,r)| \,\,\bigg|\,\, h(0,r) =h_0(r)\right\}}
\end{align*}
is finite.

\begin{lem}\label{l2}
For any $h_1,\,h_2\in X$ satisfying
\begin{align*}
\|h_1\|_X \leq x, \quad \|h_2\|_X\leq x
\end{align*}
for some $x>0$, there exists $F_2(x)\in[0,\frac{1}{2}]$ such that
\beQ
\|\mathfrak{F}(h_1)-\mathfrak{F}(h_2)\|_Y\leq F_2(x)\|h_1-h_2\|_Y.
\eeQ
\end{lem}
\pf Denote
\beQ
\mathfrak{H}=\mathfrak{F}(h_1)-\mathfrak{F}(h_2), \quad D_1=\frac{\pt}{\pt u}-\frac{q_1}{2}\frac{\pt}{\pt r}.
\eeQ
It is known that $\mathfrak{H}$ satisfies the following equation by differentiating $\mathfrak{H}$ with respect to $D_1$ (cf. (4.9) in \cite{WFAG})
\beq
D_1\mathfrak{H}=f_3\mathfrak{H}+f_4,\label{m-3}
\eeq
where
\begin{align}
f_3 =\frac{1}{2r}(g_1-q_1)+4\pi r g_2V(\bar h_1), \quad f_4 =\sum\limits_{i=1}^{12}{B_i},  \label{m-4}
\end{align}
and $B_1, \cdots, B_{12}$ are given as follows.
\begin{align*}
B_1&=\frac{1}{2}(q_1-q_2)\mathfrak{G}_2,\\
B_2&=-\frac{1}{2r}(g_1-q_1)(\bar h_1-\bar h_2),\\
B_3&=\frac{1}{2r}(g_1-q_1-g_2+q_2)\mathfrak{F}_2,\\
B_4&=-\frac{1}{2r}(g_1-q_1-g_2+q_2)\bar h_2,\\
B_5&=4\pi r(g_1-g_2)\mathfrak{F}_1V(\bar h_1),\\
B_6&=-4\pi r(g_1-g_2)\bar h_1V(\bar h_1),\\
B_7&=4\pi rg_2\left[V(\bar h_1)-V(\bar h_2)\right]\mathfrak{F}_2,\\
B_8&=-4\pi rg_2(\bar h_1-\bar h_2)V(\bar h_2),\\
B_9&=-4\pi rg_2\bar h_1\left[V(\bar h_1)-V(\bar h_2)\right],\\
B_{10}&=\frac{r}{2}(g_1-g_2)\bar h_1\frac{\pt ^2V(\bar h_1)}{\pt \bar h^2},\\
B_{11}&=\frac{r}{2}g_2(\bar h_1-\bar h_2)\frac{\pt ^2V(\bar h_1)}{\pt \bar h^2},\\
B_{12}&=\frac{r}{2}g_2\bar h_1\left[\frac{\pt ^2V(\bar h_1)}{\pt \bar h^2}-\frac{\pt ^2V(\bar h_2)}{\pt \bar h^2}\right].
\end{align*}

Now we estimate $f_3$. Similar to (\ref{l-2a}), we have
\begin{align}
|\bar h_1-\bar h_2|\leq \frac{2\|h_1-h_2\|_Y}{\ep\uu^{\ep}\ur}.\label{m-6}
\end{align}
Therefore
\begin{align}
|(h_1-h_2)-(\bar h_1-\bar h_2)| \leq 2 |\bar h_1-\bar h_2| \leq \frac{4\|h_1-h_2\|_Y}{\ep\uu^{\ep}\ur}. \label{m-7}
\end{align}
Similar to (\ref{l-3}), we have
\beq
|h_1+h_2-(\bar h_1+\bar h_2)|\leq \frac{2cxr}{\uu^{\ep-1}\ur^2}.\label{m-8}
\eeq
Multiplying (\ref{m-7}) and (\ref{m-8}), we obtain
\beQ
|(h_1-\bar h_1)^2-(h_2-\bar h_2)^2|\leq \frac{8crx \|h_1-h_2\|_Y}{\ep\uu^{2\ep-1}\ur^3}.%\label{m-9}
\eeQ
Thus,
\beq
\begin{aligned}
|g_1-g_2|&\leq 4\pi\int_{r}^{\infty}{\Big|(h_1-\bar h_1)^2-(h_2-\bar h_2)^2\Big|\frac{ds}{s}}\\
&\leq \frac{16\pi c x \|h_1-h_2\|_Y}{\ep\uu^{2\ep-1}\ur^2}.\label{m-10}
\end{aligned}
\eeq
This yields
\beq
|\bar g_1-\bar g_2|\leq \frac{1}{r}\int_{0}^{r}{|g_1-g_2|ds}\leq \frac{16\pi c x \|h_1-h_2\|_Y}{\ep\uu^{2\ep}\ur}.\label{m-11}
\eeq
On the other hand,
\beq
\begin{aligned}
|g_1-g_2-(\bar g_1-\bar g_2)|
\leq & \frac{1}{r}\int_{0}^{r}{\int_{r'}^{r}{\left|\frac{\pt(g_1-g_2)}{\pt s}\right|ds}dr'}\\
\leq & \frac{4\pi}{r}\int_{0}^{r}{\int_{r'}^{r}{\Big[|g_2||(h_2-\bar h_2)^2-(h_1-\bar h_1)^2|}}\\
&+|g_2-g_1||h_1-\bar h_1|^2 \Big]\frac{ds}{s}dr'\\
\leq &\frac{4\pi (4cx+8\pi c^3x^3)r}{\ep\uu^{2\ep}\ur^2}\|h_1-h_2\|_Y.\label{m-12}
\end{aligned}
\eeq
Using (\ref{l-2a}), we have, for any $t\in [0,1]$,
\begin{align*}
\left| t\bar h_1+(1-t)\bar h_2 \right|^p\leq \frac{2^px^p}{\ep^p\uu^{p\ep}\ur^p}.
\end{align*}
Using (\ref{m-6}) and the following identity
\begin{align*}
\bar h_1^{p+1}-\bar h_2^{p+1}=(p+1)(\bar h_1-\bar h_2)\int_{0}^{1}{\left[t\bar h_1+(1-t)\bar h_2\right]^pdt},
\end{align*}
we obtain
\begin{align*}
|\bar h_1^{p+1}-\bar h_2^{p+1}|\leq \frac{(p+1)2^{p+1}x^{p}}{\ep^{p+1}\uu^{(p+1)\ep}\ur^{p+1}}\|h_1-h_2\|_Y.
\end{align*}
Using \eqref{l-7}, we obtain
\beq
\begin{aligned}
& \bigg|\frac{8\pi}{r}\int_{0}^{r}{gs^2\left[V(\bar h_1)-V(\bar h_2)\right]ds}\bigg|\\
\leq & \frac{8\pi}{r}\int_{0}^{r}{\frac{(p+1)2^{p+1}K_0x^p\|h_1-h_2\|_Y}{\ep^{p+1}\uu^{(p+1)\ep}}\frac{s^2ds}{(1+u+s)^{p+1}}}\\
\leq &\frac{8(p+1)\pi C_1 x^pr^2}{\uu^{(\omega+1)\ep+1}\ur^{\omega}}\|h_1-h_2\|_Y.  \label{m-14}
\end{aligned}
\eeq
Let $C_{11}=\max{\left\{\frac{16\pi c}{\ep},8(p+1)\pi C_1\right\}}$, $C_{12}=\max{\left\{\frac{16\pi^2 c^3}{\ep},\,\,4(p+1)\pi C_1\right\}}$.
Using (\ref{m-11}), (\ref{m-12}), and (\ref{m-14}), we obtain
\beq
\begin{aligned}
|q_1-q_2|&\leq |\bar g_1-\bar g_2|+\frac{8\pi}{r}\int_{0}^{r}{
gs^2\left|V(\bar h_1)-V(\bar h_2)\right|ds}\\
&\leq \frac{C_{11}(x+x^p)r}{\uu^{2\ep+1}\ur}\|h_1-h_2\|_Y,
\end{aligned}
\label{m-15}
\eeq
and
\beq
\begin{aligned}
\frac{1}{2r}|g_1-g_2-(q_1-q_2)|
\leq &\frac{1}{2r}|g_1-g_2-(\bar g_1-\bar g_2)|\\
&+\frac{4\pi}{r^2}   \left|\int_{0}^{r}{gs^2\left[V(\bar h_1)-V(\bar h_2)\right]ds}\right|\\
\leq &\frac{C_{12}(x+x^3+x^p)}{\uu^{2\ep}\ur^{2}}\|h_1-h_2\|_Y.\label{m-16}\\
\end{aligned}
\eeq
Therefore, we obtain
\begin{align}
\int_{0}^{u_1}{|f_3|du}\leq C_3(x^2+x^{p+1}). \label{m-17}
\end{align}

Next, we estimate $f_4$. Denote
\begin{align*}
\alpha(x)=&\frac{C_{10}C_{11}}{2}(x+x^p)(d+x^3+x^p+x^{p+2})\exp[C_{10}(x^2+x^{p+1})] P(x),\\
\beta(x)=&C_5C_{12}(x+x^3+x^p)(d+x^3+x^p+x^{p+2})\exp[C_3(x^2+x^{p+1})],\\
\gamma(x)=&\frac{2^{p+3}\pi K_0C_5x^p}{\ep^{p+1}}(d+x^3+x^p+x^{p+2})\exp[C_3(x^2+x^{p+1})],\\
\sigma(x)=&\frac{2^{p+7}K_0\pi^2cC_5x^{p+2}}{\ep^{p+2}}(d+x^3+x^p+x^{p+2})\exp[C_3(x^2+x^{p+1})].
\end{align*}
Using (\ref{l1-a}), (\ref{l1-b}), (\ref{l-8}), (\ref{m-6}), (\ref{m-10}), (\ref{m-15}), and (\ref{m-16}), we have
\begin{align*}
|B_1| \leq & \frac{1}{2}\frac{C_{10}(d+x^3+x^p+x^{p+2})\exp\left[C_{10}(x^2+x^{p+1})\right] P(x)}{\ur^{1+\ep}}\\
           & \times \frac{C_{11}(x+x^p)r}{\uu^{2\ep+1}\ur}\|h_1-h_2\|_Y\\
      \leq & \frac{\alpha(x)}{\uu^{(\omega-1)\ep+\omega-2}\ur^{1+\ep}}\|h_1-h_2\|_Y,\\
|B_2| \leq & \frac{1}{2r}\frac{C_2r^2(x^2+x^{p+1})}{\uu^{2\ep+2-\omega}\ur^{\omega}}\frac{2\|h_1-h_2\|_Y}{\ep\uu^{\ep}\ur}\\
      \leq & \frac{C_2(x^2+x^{p+1})}{\ep\uu^{(\omega-1)\ep+\omega-2}\ur^{1+\ep}}\|h_1-h_2\|_Y,\\
|B_3| \leq & \frac{C_{12}(x+x^3+x^p)}{\uu^{2\ep}\ur^{2}}\|h_1-h_2\|_Y\\
           & \times\frac{C_5(d+x^3+x^p+x^{p+2})\exp\left[C_3(x^2+x^{p+1})\right]}{\ur^{1+\ep}}\\
      \leq & \frac{\beta(x)}{\uu^{(\omega-1)\ep+\omega-2}\ur^{1+\ep}}\|h_1-h_2\|_Y,\\
|B_4| \leq & \frac{C_{12}(x+x^3+x^p)\|h_1-h_2\|_Y}{\uu^{2\ep}\ur^{2}}\frac{2x}{\ep\uu^{\ep}\ur}\\
      \leq & \frac{2C_{12}(x^2+x^4+x^{p+1})}{\ep\uu^{(\omega-1)\ep+\omega-2}\ur^{1+\ep}}\|h_1-h_2\|_Y,\\
|B_5| \leq & \frac{64\pi^2 cxr\|h_1-h_2\|_Y}{\ep\uu^{2\ep-1}\ur^2}\frac{2^{p+1}x^{p+1}K_0}{\ep^{p+1}(1+u)^{(p+1)\ep}(1+u+r)^{p+1}}\\
           & \times\frac{C_5(d+x^3+x^p+x^{p+2})\exp\left[C_3(x^2+x^{p+1})\right]}{\ur^{1+\ep}}\\
      \leq & \frac{\sigma(x)}{\uu^{(\omega-1)\ep+\omega-2}\ur^{1+\ep}}\|h_1-h_2\|_Y,\\
|B_6| \leq & \frac{4\pi r\cdot 16\pi cx\|h_1-h_2\|_Y}{\ep\uu^{2\ep-1}\ur^2}\frac{2x}{\ep\uu^{\ep}\ur}\\
           & \times\frac{2^{p+1}x^{p+1}K_0}{\ep^{p+1}\uu^{(p+1)\ep}\ur^{p+1}}\\
      \leq & \frac{2^{p+8}\pi^2cK_0x^{p+3}}{\ep^{p+3}\uu^{(\omega-1)\ep+\omega-2}\ur^{1+\ep}}\|h_1-h_2\|_Y,\\
|B_7| \leq & \frac{4\pi r 2^{p+1}x^pK_0\|h_1-h_2\|_Y}{\ep^{p+1}\uu^{(p+1)\ep}\ur^{p+1}}\\
           & \times\frac{C_5(d+x^3+x^p+x^{p+2})\exp\left[C_3(x^2+x^{p+1})\right]}{\ur^{1+\ep}}\\
      \leq & \frac{\gamma(x)}{\uu^{(\omega-1)\ep+\omega-2}\ur^{1+\ep}}\|h_1-h_2\|_Y,\\
|B_8| \leq & 4\pi r\frac{2\|h_1-h_2\|_Y}{\ep\uu^{\ep}\ur}\frac{2^{p+1}x^{p+1}K_0}{\ep^{p+1}\uu^{(p+1)\ep}\ur^{p+1}}\\
      \leq & \frac{2^{p+4}\pi K_0x^{p+1}}{\ep^{p+2}\uu^{(\omega-1)\ep+\omega-2}\ur^{1+\ep}}\|h_1-h_2\|_Y,\\
|B_9| \leq & 4\pi r\frac{2x}{\ep\uu^{\ep}\ur}\frac{2^{p+1}x^pK_0\|h_1-h_2\|_Y}{\ep^{p+1}\uu^{(p+1)\ep}\ur^{p+1}}\\
      \leq & \frac{2^{p+4}\pi K_0x^{p+1}}{\ep^{p+2}\uu^{(\omega-1)\ep+\omega-2}\ur^{1+\ep}}\|h_1-h_2\|_Y,\\
|B_{10}|\leq & \frac{r}{2}\frac{16\pi cx\|h_1-h_2\|_Y}{\ep\uu^{2\ep-1}\ur^2}\frac{2x}{\ep\uu^{\ep}\ur}\\
             & \times\frac{2^{p-1}x^{p-1}K_0}{\ep^{p-1}\uu^{(p-1)\ep}\ur^{p-1}}\\
        \leq & \frac{2^{p+3}\pi cK_0x^{p+1}}{\ep^{p+1}\uu^{(\omega-1)\ep+\omega-2}\ur^{1+\ep}}\|h_1-h_2\|_Y,\\
|B_{11}| \leq & \frac{r}{2}\frac{2\|h_1-h_2\|_Y}{\ep\uu^{\ep}\ur}\frac{2^{p-1}x^{p-1}K_0}{\ep^{p-1}\uu^{(p-1)\ep}\ur^{p-1}}\\
         \leq & \frac{2^{p-1}x^{p-1}K_0}{\ep^{p}\uu^{(\omega-1)\ep+\omega-2}\ur^{1+\ep}}\|h_1-h_2\|_Y,\\
|B_{12}| \leq & \frac{r}{2}\frac{2x}{\ep\uu^{\ep}\ur}\frac{2^{p-1}x^{p-2}K_0\|h_1-h_2\|_Y}{\ep^{p-1}\uu^{(p-1)\ep}\ur^{p-1}}\\
         \leq & \frac{2^{p-1}x^{p-1}K_0}{\ep^{p}\uu^{(\omega-1)\ep+\omega-2}\ur^{1+\ep}}\|h_1-h_2\|_Y.
\end{align*}
Let $C_{13}=\max{\left\{\frac{2^pK_0}{\ep^p},\frac{C_2+2C_3}{\ep}+\frac{2^{p+3}\pi K_0(4+c\ep)}{\ep^{p+2}},\frac{2^{p+8}\pi^2cK_0}{\ep^{p+3}}\right\}}$. Denote
\begin{align*}
F_3(x)=\alpha(x)+\beta(x)+\gamma(x)+\sigma(x)+x^2+x^4+x^{p-1}+x^{p+1}+x^{p+3}.
\end{align*}
The above estimates yield
\begin{align}
|f_4|\leq \frac{C_{13}F_3(x)}{\uu^{(\omega-1)\ep+\omega-2}\ur^{1+\ep}}\|h_1-h_2\|_Y. \label{f4}
\end{align}

Using (\ref{m-3}), (\ref{m-4}), (\ref{m-17}) and \eqref{f4}, we obtain
\begin{align*}
|\mathfrak{H}(u_1,r_1)|&\leq \int_{0}^{u_1}{\exp\left(\int_{u}^{u_1}{|f_3|_{\chi}du'}\right)|f_4|_{\chi}du}\\
&\leq \frac{C_{13}F_3(x)\exp[C_3(x^2+x^{p+1})]}{[(\omega-1)\ep+\omega-3]k^{1+\ep}(1+u_1+r_1)^{1+\ep}}\|h_1-h_2\|_Y.
\end{align*}
This implies that
\beQ
\|\mathfrak{H}\|_{Y}\leq F_2(x)\|h_1-h_2\|_Y,
\eeQ
where
\begin{align*}
F_2(x)=\frac{C_{13}\exp[C_{12}(x^2+x^{p+1})]}{[(\omega-1)\ep+\omega-3]k^{1+\ep}}F_3(x).
\end{align*}
Clearly,
\beQ
F_2(0)=0,\quad F_2'(x)\geq 0.
\eeQ
Therefore $F_2(x)$  is monotonically increasing on $[0,\infty)$. Thus, there exists an $x_2>0$ such that for any $x\in (0,x_2]$,
\beQ
0\leq F_2(x)\leq \frac{1}{2}.
\eeQ
Hence, the mapping $h\mapsto \mathfrak{F}(h)$ contracts in $Y$ for $\|h\|_X\leq x_2$.
\qed

\mysection{Global existence and uniqueness}
\ls

In this section, we prove the main theorem.

{\em Proof of Theorem \ref{M}.} Take
\beQ
\tilde x=\min{\{x_0,x_2,1\}},\,\,\delta=\max\limits_{x\in[0,\tilde x] }{F_1(x)}.
\eeQ
If $d<\delta$, there exists $x\in (0,\tilde x]$ such that $d\leq F_1(x)$. Then Lemma \ref {l1} and Lemma \ref{l2} imply that $h\mapsto \mathfrak{F}(h)$ is a contraction mapping in $X$. Banach's fixed point theorem shows that there exists a unique fixed point $h\in X$ such that
\begin{align}
\mathfrak{F}(h)=h. \label{fixed point}
\end{align}
Using the explicit repression of $\mathfrak{F}(h)$ given by \eqref{int-f} and taking the $D$-derivative on both sides of \eqref{fixed point} for $r>0$, we find that $h$ is the unique solution to (\ref{e-k-g-2}) for $r>0$. But we still need to show that the solution can extend to $r=0$. This can be done in the spirit of \cite{C1, LZ} by showing that $\frac{\partial h}{\partial r}$ is uniformly continuous with respect to $r$ for $r \geq 0$.

Indeed, let $\chi_1(u;r_1)$ and $\chi_2(u;r_2)$ be two characteristics through the line $u=u_1$ at $r=r_1 \geq 0$ and $r=r_2 \geq r_1$ respectively. Define
\beQ
\psi(u)=\frac{\pt\mathfrak{F}}{\pt r}\big(u,\chi_1(u;r_1)\big)-\frac{\pt\mathfrak{F}}{\pt r}\big(u,\chi_2(u;r_2)\big).
\eeQ
Differentiating $\psi(u)$, we have
\beq
\begin{aligned}
\psi'(u)=&f_1\big(u,\chi_1(u;r_1)\big)\psi(u)\\
          &-\bigg(f_1 \big(u,\chi_2(u;r_2) \big)-f_1 \big(u,\chi_1(u;r_1)\big)\bigg)
\frac{\pt \mathfrak{F}}{\pt r} \big(u,\chi_2(u;r_2)\big)\\
&-f_2\big(u,\chi_2(u;r_2)\big)+f_2\big(u,\chi_1(u;r_1)\big),
\end{aligned}\label{p}
\eeq
where $f_1$, $f_2$ are given in \eqref{l-19}. Note that
\begin{align*}
\frac{\pt^2 q}{\pt r^2}=&\frac{\pt^2\bar g}{\pt r^2}+\frac{16\pi }{r^3}\int_{0}^{r}{gs^2V(\bar h)ds}\\
                        &+32\pi^2 g(h-\bar h)^2V(\bar h)+8\pi g(h-\bar h)\frac{\pt V(\bar h)}{\pt \bar h},\\
\frac{\pt f_1}{\pt r}=&\frac{1}{2}\frac{\pt^2 q}{\pt r^2}-\frac{g-q}{2r^2}+\frac{1}{2r}\frac{\pt }{\pt r}(g-q)\\
&+4\pi\frac{\pt g}{\pt r}rV(\bar h)+4\pi gV(\bar h)+4\pi gr\frac{\pt V(\bar h)}{\pt \bar h}\frac{\pt \bar h}{\pt r},
\end{align*}
we obtain
\begin{align*}
\left|\frac{\pt f_1}{\pt r}\right|\leq & \frac{C(x^2+x^{p+1}+x^{p+3})}{\uu^{\mu}},\\
\left|\chi_1(u;r_1)-\chi_2(u;r_2)\right|\leq & (r_2-r_1)\exp{\left[\frac{C(x^2+x^{p+1})}{4\ep}\right]}.
\end{align*}
Therefore
\begin{align*}
\bigg|\bigg(f_1 \big(u,\chi_2(u;r_2)\big) &-f_1\big(u,\chi_1(u;r_1)\big)\bigg)
\frac{\pt \mathfrak{F}}{\pt r}\big(u,\chi_2(u;r_2)\big)\bigg|\\
& \leq \frac{A_1(d,x)(r_2-r_1)}{\uu^{\mu}},
\end{align*}
where $A_1(d,x)$ is independent of $u$ and $r$. Since $f_2$ is continuous and satisfies
\begin{align*}
\left|(1+u)^{\mu}f_2 \right|\leq \frac{A_2(d,x)}{\ur}
\end{align*}
where $A_2(d,x)$ is independent of $u$ and $r$, we have
\begin{align*}
\lim _{r \rightarrow \infty} (1+u)^{\mu}f_2 =0.
\end{align*}
Thus, $(1+u)^{\mu}f_2$ is uniformly continuous. For any $\eta>0$, there exists $s_1>0$ such that, if
\begin{align*}
|\chi_2(u;r_2)-\chi_1(u;r_1)|\leq s_1,
\end{align*}
then
\begin{align*}
\left|f_2 \big(u,\chi_2(u;r_2)\big)-f_2\big(u,\chi_1(u;r_1)\big)\right|\leq \frac{(\mu-1)\eta\exp{[-C(x^2+x^{p+1})]}}{3\uu^{\mu}}.
\end{align*}
Denote $k'=\exp{\left[\frac{C(x^2+x^{p+1})}{4\ep}\right]}$. We have
\begin{align*}
|r_2-r_1|\leq \frac{s_1}{k'} \Longrightarrow |\chi_2(u;r_2)-\chi_1(u;r_1)|\leq s_1.
\end{align*}
Since
\begin{align*}
\int_{0}^{u_1}{|f_1|_{\chi}du}\leq C(x^2+x^{p+1}),
\end{align*}
there exists $s_2>0$ such that, for $|r_2-r_1|\leq s_2$,
\begin{align*}
|\psi(0)|=\bigg| \frac{\pt h}{\pt r}\big(0,\chi_1(0;r_1)\big)-\frac{\pt h}{\pt r}\big(0,\chi_2(0;r_2)\big)\bigg|
        \leq \frac{\eta\exp{[-C(x^2+x^{p+1})]}}{3}.
\end{align*}
Taking
\begin{align*}
s=\min{\left\{\frac{s_1}{k'},s_2,\frac{(\mu-1)\eta\exp{[-C(x^2+x^{p+1})]}}{3A_1(d,x)}\right\}}
\end{align*}
and integrating (\ref{p}), we have
\beQ
r_2-r_1\leq s \Longrightarrow |\psi(u_1)|=\left|\frac{\pt \mathfrak{F}}{\pt r}(u_1,r_1)-\frac{\pt \mathfrak{F}}{\pt r}(u_1,r_2)\right|\leq \eta.
\eeQ
This implies $\frac{\pt \mathfrak{F}}{\pt r}$ is uniformly continuous with respect to $r$. Therefore $h$ is the unique solution to equation (\ref{e-k-g-2}) for $r\geq 0$. The decaying estimates can be obtained from Lemma \ref{l1} directly.

Using the first equation in (\ref{e-k-g-2}), we can show that $g$ is monotonically increasing and 
\begin{align*}
0<k<\bar{g}\leq g\leq 1. 
\end{align*}
Moreover, $m(u,r)=\frac{r}{2}\Big(1-\frac{\bar{g}}{g}\Big)$ satisfies
\begin{align*}
\frac{\pt m}{\pt r}=2\pi\frac{\bar g}{g}(h-\bar h)^2
\end{align*}
(cf. (4.4) in \cite{C1}). As $m(u,0)=0$, it gives
\begin{align*}
m(u,r)=2\pi\int_{0}^{r}{\frac{\bar g}{g}(h-\bar h)^2ds}.
\end{align*}
As \eqref{l-2a} implies that
\begin{align*}
|(h-\bar h)(u,r)|\leq |h(u,r)|+|\bar h(u,r)|\leq \frac{(2+\ep)C}{\ep\uu^{\ep}\ur}
\end{align*}
for some constant $C>0$, we obtain
\begin{align*}
m(u,r)\leq \frac{2(2+\ep)^2C^2\pi}{\ep^2}\int_{0}^{r}{\frac{ds}{\uu^{2\ep}(1+u+s)^2}}
\leq \frac{2(2+\ep)^2C^2\pi}{\ep^2\uu^{2\ep+1}}.
\end{align*}
Thus
\begin{align*}
\frac{r}{2}\Big|1-q\Big|
&\leq \frac{r}{2}\Big(1-\bar g\Big)+\frac{r}{2}\frac{8\pi}{r}\int_{0}^{r}{gs^2|V(\bar h)|ds}\\
&= \frac{r}{2}\Big(1-g\Big)+m(u,r) g+4\pi \int_{0}^{r}{gs^2|V(\bar h)|ds}.
\end{align*}
Note that (\ref{l-3}) implies
\beQ
\lim\limits_{r\rw\infty}{\frac{r}{2}\Big(1-g\Big)}=0.
\eeQ
Thus, using \eqref{l-7}, we obtain
\beQ
|M_B(u)|\leq \frac{2(2+\ep)^2C^2\pi}{\ep^2\uu^{2\ep+1}}+
\frac{2^{p+2}\pi K_0x^{p+1}}{3\ep^{p+1}\uu^{(\omega+1)\ep+1}}.
\eeQ
This shows the final Bondi mass vanishes. As $q>0$ by \eqref{qq}, we have
\begin{align*}
M(u) \leq M_B(u).
\end{align*}
This shows the final Bondi-Christodoulou mass also vanishes.

Finally, from (\ref{V}), (\ref{l-3}), and (\ref{l-8}), we know that there is some constant $C_0 >0$ such that
\begin{align*}
\left|\frac{\pt h}{\pt u}\right|\leq C_0\Longrightarrow \left|\frac{\pt \bar h}{\pt u}\right|\leq \frac{1}{r}\int_{0}^{r}{\left|\frac{\pt h}{\pt u}\right|dr}\leq C_0.
\end{align*}
Thus
\begin{align*}
\left|\frac{\pt g}{\pt u}\right|\leq 8\pi\int_{r}^{\infty}{|h-\bar h|\cdot \left|\frac{\pt h}{\pt u}-\frac{\pt \bar h}{\pt u}\right|\frac{dr}{r}}\leq 16\pi c C_0 \Longrightarrow \left|\frac{\pt\bar g}{\pt u}\right|\leq 16\pi c C_0.
\end{align*}
Therefore
\begin{align*}
\left|\frac{\pt q}{\pt u}\right| \leq \left|\frac{\pt\bar g}{\pt u}\right|+\frac{8\pi}{r}\int_{0}^{r}{gs^2\left|\frac{\pt V(\bar h)}{\pt \bar h}\right|\cdot\left|\frac{\pt \bar h}{\pt u}\right|ds}
\leq 16\pi cC_0+\frac{8\pi C_0K_0}{\ep^p}.
\end{align*}
These together with (\ref{l-7}), (\ref{l-23}), and (\ref{l-22}) give
\beQ
0<k\leq q \leq 1+8\pi C_1.
\eeQ
Moreover, $\frac{\pt h}{\pt u}$, $\frac{\pt \bar h}{\pt u}$, $\frac{\pt g}{\pt u}$, $\frac{\pt g}{\pt r}$, $\frac{\pt q}{\pt u}$, $\frac{\pt q}{\pt r}$
are all uniformly bounded. Using the method in \cite{LZ,LOY}, we conclude that the corresponding spactime is future causally geodesically complete. \qed

\bigskip

{\footnotesize {\it Acknowledgement} The authors are grateful to the referees for many valuable suggestions to improve both the main result and the presentation of the paper. The work is supported by the National Natural Science Foundation of China 12301072, 12326602, the special foundation for Junwu and Guangxi Ba Gui Scholars.}


\begin{thebibliography}{99}
\bibitem{BBM} H. Bondi, M. van ber Burg, A. Metzner, Gravitational waves in general relativity. VII. Waves from axi-symmetric isolated systems. Proc. Roy. Soc. London Ser. A 26 (1962).
\bibitem{C01} D. Chae, Global existence of spherically symmetric solutions to the coupled Einstein and nonlinear Klein-Gordon system. Class. Quantum Gravity 18, 4589-4605 (2001).
\bibitem{C02} D. Chae, Global existence of solutions to the coupled Einstein and Maxwell-Higgs system in the spherical symmetry. Ann. Henri Poincar\'{e} 4, 35-62 (2003).
\bibitem{C1} D. Christodoulou, The problem of a self-gravitating scalar field. Commun. Math. Phys. 105, 337-361 (1986).
\bibitem{C2} D. Christodoulou, Global existence of generalized solutions of the spherically symmetric Einstein-scalar equations in the large. Commun. Math. Phys. 106, 587-621 (1986).
\bibitem{C3} D. Christodoulou, The structure and uniqueness of generalized solutions of the spherically symmetric Einstein-scalar equations. Commun. Math. Phys. 109, 591-611 (1987).
\bibitem{C4} D. Christodoulou, Bounded variation solutions of the spherically symmetric Einstein-scalar field equations. Comm. Pure Appl. Math. 46(8), 1131-1220 (1993).
\bibitem{FWY} A. Fang, Q. Wang, S. Yang, Global solution for massive Maxwell-Klein-Gordon equations with large Maxwell field. Ann. PDE. 7:3 (2021).
\bibitem{K} S. Klainerman, Global existence of small amplitude solutions to nonlinear Klein-Gordon equations in four space-time dimensions. Comm. Pure Appl. Math. 38(5), 631-641 (1985).
\bibitem{KWY} S. Klainerman, Q. Wang, S. Yang, Global Solution for Massive Maxwell-Klein-Gordon Equations. Comm. Pure Appl. Math. 73, 63-109 (2020).
\bibitem{ML} P. LeFloch, Y. Ma, The global nonlinear stability of Minkowski space for self-gravitating massive fields. Commun. Math. Phys. 346, 603-665 (2016).
\bibitem{LR} H. Lindblad, I. Rodnianski, Global stability of Minkowski space-time in harmonic gauge. Ann. Math. 171(3), 1401-1477 (2010).
\bibitem{LZ} C. Liu, X. Zhang, Spherically symmetric Einstein-scalar-field equations for wave-like decaying null infinity. Adv. Math. 409, 108642 (2022).
\bibitem{LO} J. Luk, S-J. Oh, Quantitative decay rates for dispersive solutions to the Einstein-scalar field system in spherical symmetry. Ann. PDE. 8(7), 1603-1674 (2015).
\bibitem{LO1} J. Luk, S-J. Oh, Global nonlinear stability of large dispersive solutions to the Einstein equations. Ann. Henri Poinc\'{a}re, 23, 2391-2521 (2022).
\bibitem{LOY} J. Luk, S-J. Oh, S. Yang, Solutions to the Einstein-scalar-field system in spherical symmetry with large bounded variation norms. Ann. PDE. 4(3), 1-63 (2018).
\bibitem{W} Q. Wang, An intrinsic hyperboloid approach for Einstein Klein-Gordon equations. J. Differ. Geom. 115(1), 27-109 (2020).
\bibitem{WFAG} M. Wijayanto, E. Fadhila, F. Akbar, B. Gunara, Global existence of classical static solutions of four dimensional Einstein-Klein-Gordon system. Gen. Relat. Gravit. 55:19 (2023).


\end{thebibliography}
\end{document}